\documentclass[12pt]{iopart}

\usepackage{bm}
\usepackage{graphicx}
\usepackage{epsfig}
\usepackage{dcolumn}
\usepackage{bm}

\usepackage{soul}
\usepackage{xcolor}

\usepackage{url}

\begin{document}

\title{The Fourier signatures of memristive hysteresis}

\author{Yuriy V. Pershin$^1$, Chih-Chun Chien$^2$, Massimiliano Di Ventra$^3$}

\address{$^1$Department of Physics and Astronomy, University of South Carolina, Columbia, South Carolina 29208, USA}
\address{$^2$Department of Physics, University of California, Merced, CA 95343, USA}%
\address{$^3$Department of Physics, University of California, San Diego, La Jolla, CA 92093, USA}%
\ead{pershin@physics.sc.edu}
\ead{diventra@physics.ucsd.edu}
\vspace{10pt}
\begin{indented}
\item[]December 2020
\end{indented}

\begin{abstract}
While resistors with memory, sometimes called memristive elements (such as ReRAM cells), are often studied under conditions of periodic driving, little attention has been paid to the Fourier features of their memory response (hysteresis). Here we demonstrate experimentally that the hysteresis of memristive systems can be unambiguously distinguished from the linear or non-linear response of systems without hysteresis by the values of certain Fourier series coefficients. We also show that the Fourier series convergence depends on driving conditions, and introduce a measure of hysteresis. These results may be used to quantify the memory content of resistive memories, and tune their
Fourier spectrum according to the excitation signal.
\end{abstract}

%
%
%
%
%

\section{Introduction}

Despite being a well-known and useful technique, Fourier analysis has been rarely applied to resistive memories, sometimes called
memristive devices~\cite{chua76a}. In fact, only a handful of publications have appeared in the literature that discuss the memory response (hysteresis) of
memristive elements by means of Fourier analysis~\cite{Joglekar12a,Cohen12a,Biolek14a,Hu19a}.

For instance, Joglekar and Meijome~\cite{Joglekar12a} studied the properties of Fourier harmonics within a particular memristive model.
They demonstrated that the frequency content of periodically-driven memristive systems depends on the driving frequency, and the role of higher harmonics is more important at lower excitation frequencies. Moreover, two of the present authors (YVP and MD) investigated the power conversion into the second and higher harmonics in a memristive bridge, and shown that memristive bridges may provide a more efficient power conversion compared to the diode bridges~\cite{Cohen12a}.

However, none of these studies have identified the distinctive features of the hysteresis in the Fourier series of individual memristive elements, compared to the response of devices without memory. In view of the fact that Fourier analysis is a powerful tool to characterize the response of {\it any} physical system, the question arises as to
whether the hysteresis of a memristive system showcases specific signatures in its Fourier series that distinguish it clearly from other systems, whether linear or not, without hysteresis. Not only such a study will provide an additional characterization tool to study the response of systems with memory, it may
also guide their applicative aspect by suggesting how these systems should be excited by external perturbations to modify their Fourier spectrum.

In this work we will demonstrate experimentally, using commercially available memristive devices~\cite{knowm}, that their hysteresis can be characterized by the values of certain Fourier series coefficients, compared to their memory-less counterparts (whether linear or non-linear). We also discuss the important effect the waveform of the driving signal has on the Fourier series, which can be used as a tool to tune their response.

A Fourier series is a representation of a function in the form
\begin{equation}\label{eq:1}
  f(x)=\frac{a_0}{2}+\sum\limits_{n=1}^{\infty}a_n\cos nx+\sum\limits_{n=1}^{\infty}b_n\sin nx,
\end{equation}
where a $2\pi$-periodicity is assumed, and the coefficients $a_n$ and $b_n$ are calculated using the well-known integral expressions~\cite{arfken1999mathematical}.
In their textbook~\cite{arfken1999mathematical}, Arfken and Weber argue  that there are two characteristic features of Fourier expansions~\cite{raisbeck1955order}:
\begin{description}
 \item[$\bullet$] {\it Feature 1:} For functions with {\it discontinuities}, the $n$-th coefficient decreases as $1/n$ (a relatively slow convergence).
 \item[$\bullet$] {\it Feature 2:} For {\it continuous} functions (possibly with discontinuous derivatives), the $n$-th coefficient decreases as $1/n^2$ {\it  or faster}~\footnote{The words `{\it or faster}' were added by us to take into account the well-known fact that for certain functions, e.g., the parabolic wave, the coefficient decrease is faster than $1/n^2$.}.
\end{description}
In fact, without being mentioned in the analysis of  Ref.~\cite{Joglekar12a}, these features explain the trends reported in that work.

In this paper, we focus on passive resistive devices with very small/negligible capacitive and/or inductive components in their response. It is assumed that the devices are deterministic (noise effects are negligible), and exhibit a periodic response under conditions of periodic driving. We compare two very different excitation signals: a sinusoidal driving (without dc offset) and a triangular wave voltage of given frequency $\omega$. We also define a measure of hysteresis based on the relative contribution of the Fourier components of the Fourier spectrum characteristic of the memory response, and apply this measure to experimental and model data.

\section{Experimental results}

The results reported in this paper were obtained using a conventional 100 Ohm resistor, a 1N4148 silicon diode, and a commercially available electrochemical metallization (ECM) cell from Knowm, Inc.~\cite{knowm} The latter is a memristive component, whose resistance change is caused by a voltage-induced drift of Ag atoms through a stack of chalcogenide layers with one of the Ge$_2$Se$_3$ layers doped by Cr (M+SDC Cr device). The current was recorded as a function of voltage in quasi-static measurements (with a period of the order of one minute) performed with a Keithley 236 source measure unit. Fourier series coefficients were calculated using a custom code.

 \begin{figure}[tb]
 \begin{center}
(a)\includegraphics[angle=0,width=7cm]{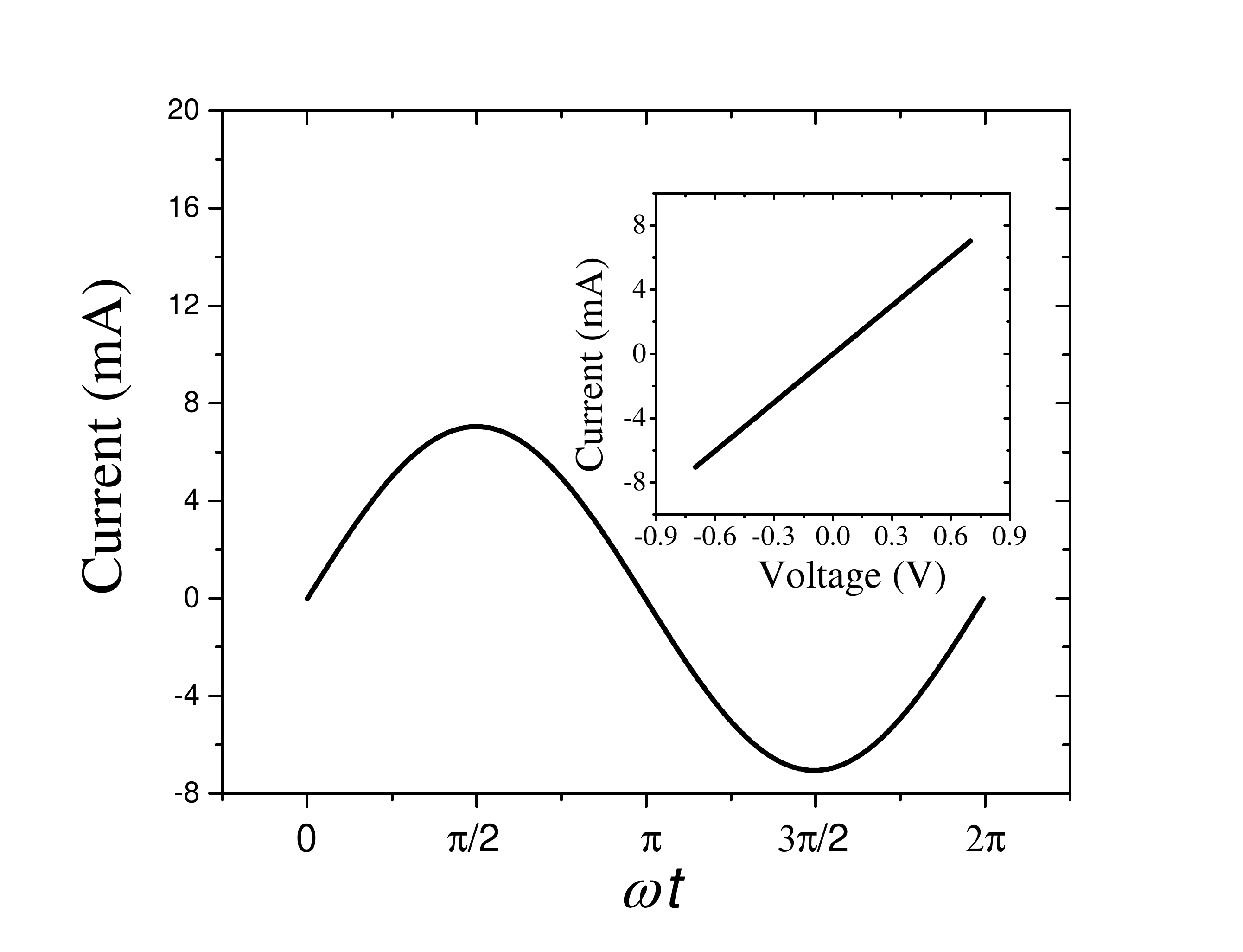}
(c)\includegraphics[angle=0,width=7cm]{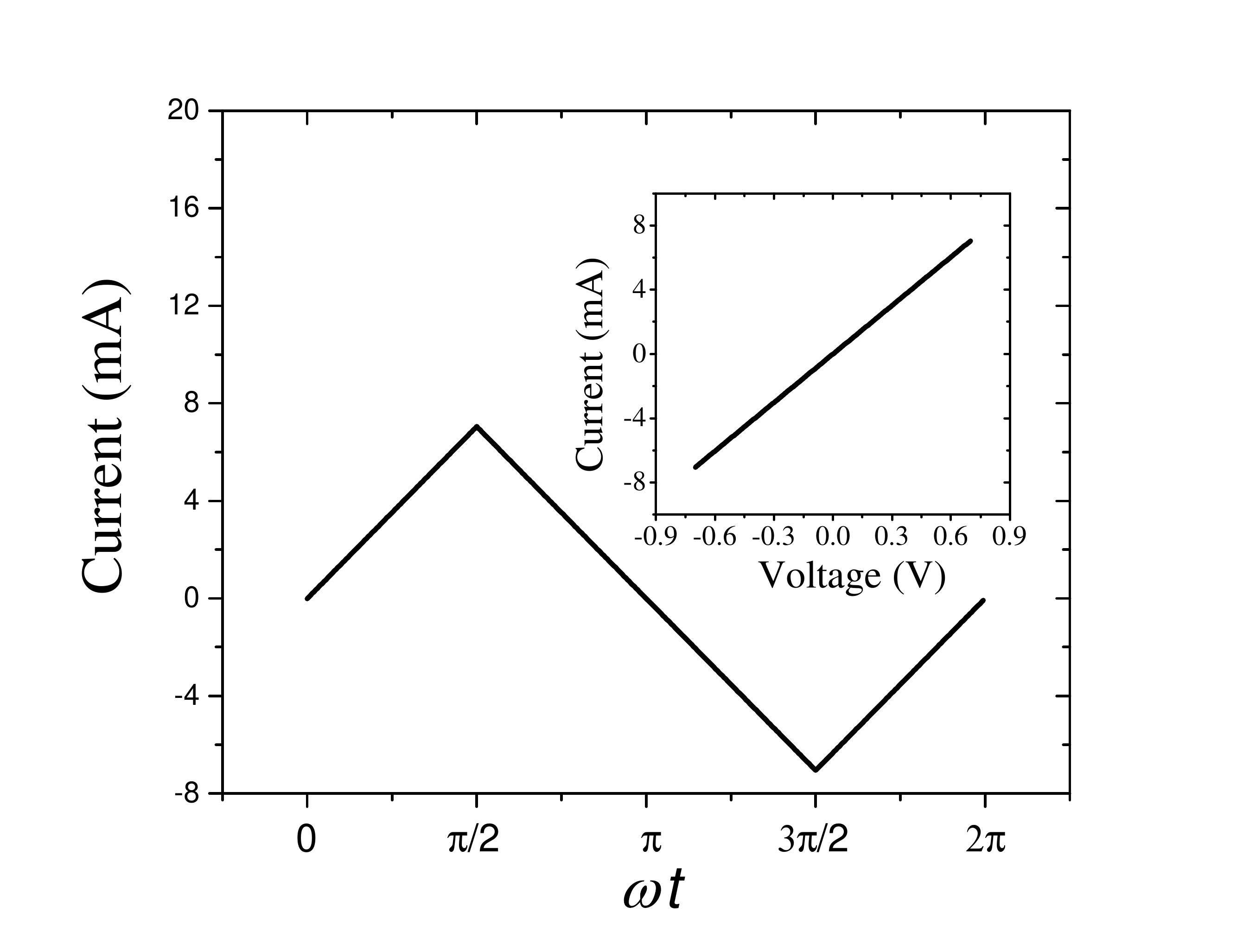}
(b)\includegraphics[angle=0,width=7cm]{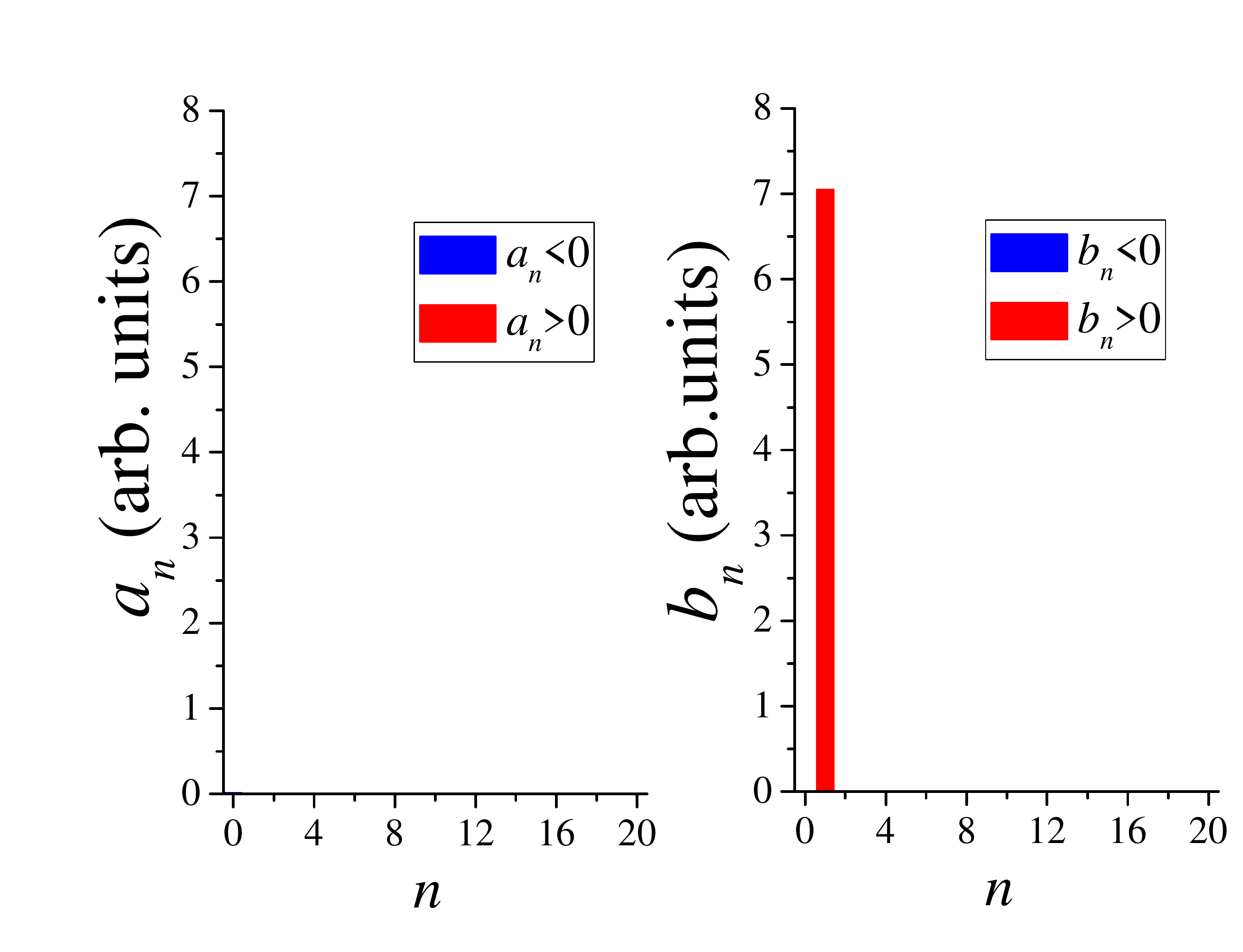}
(d)\includegraphics[angle=0,width=7cm]{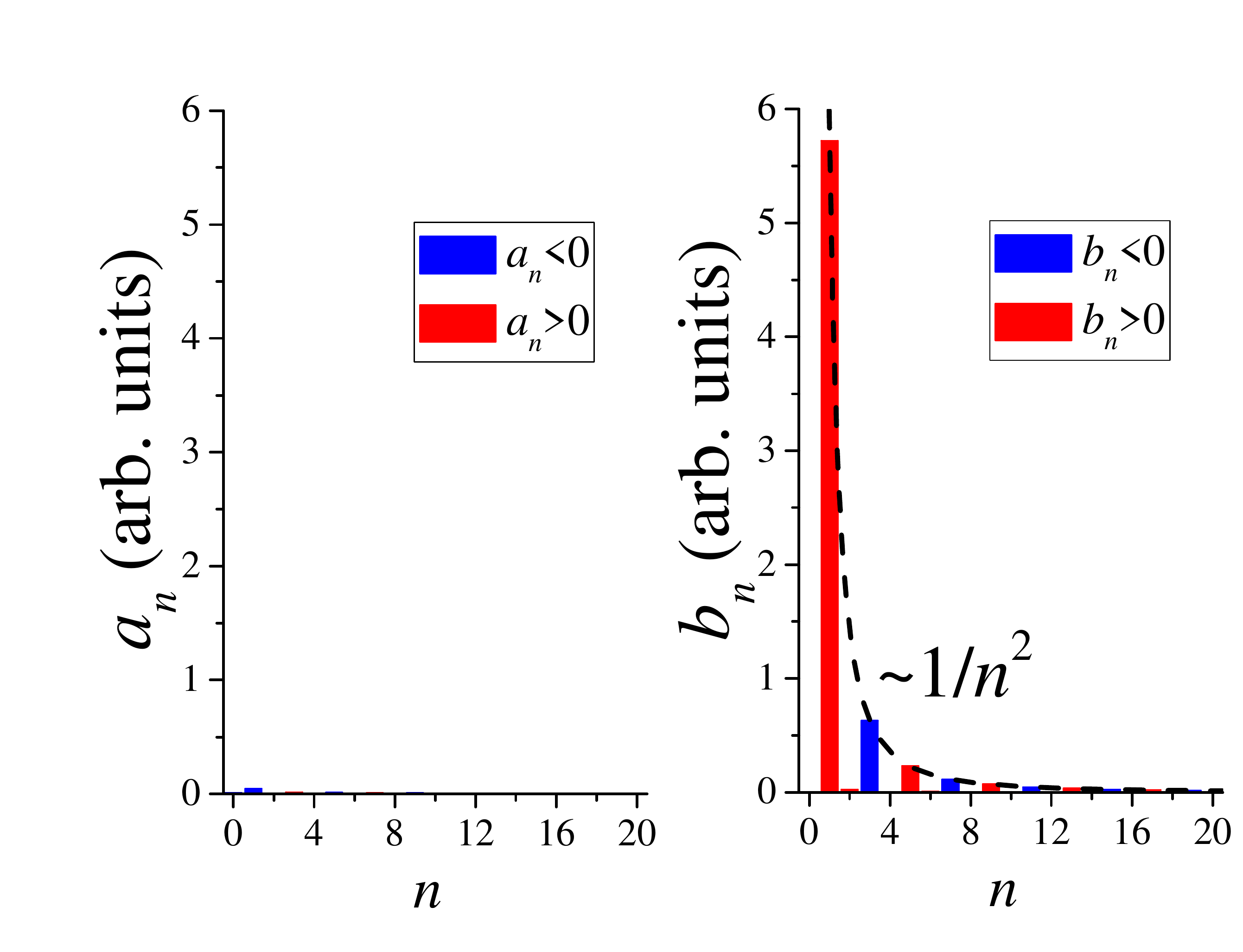}
\end{center}
\caption{A linear resistor driven by sinusoidal (left) and triangular (right) voltage waveforms. (a), (c) Current as a function of $\omega t$, and (b), (d) coefficients of Fourier series. In (d) the $1/n^2$ fit is also shown. Insets in (a) and (c): current-voltage curves. Because of the symmetry in the resistor response ($I(-V)=-I(V)$), there are no cosine terms in the Fourier series for resistors.  }\label{fig:R}
\end{figure}

 \begin{figure}[tb]
 \begin{center}
(a)\includegraphics[angle=0,width=7cm]{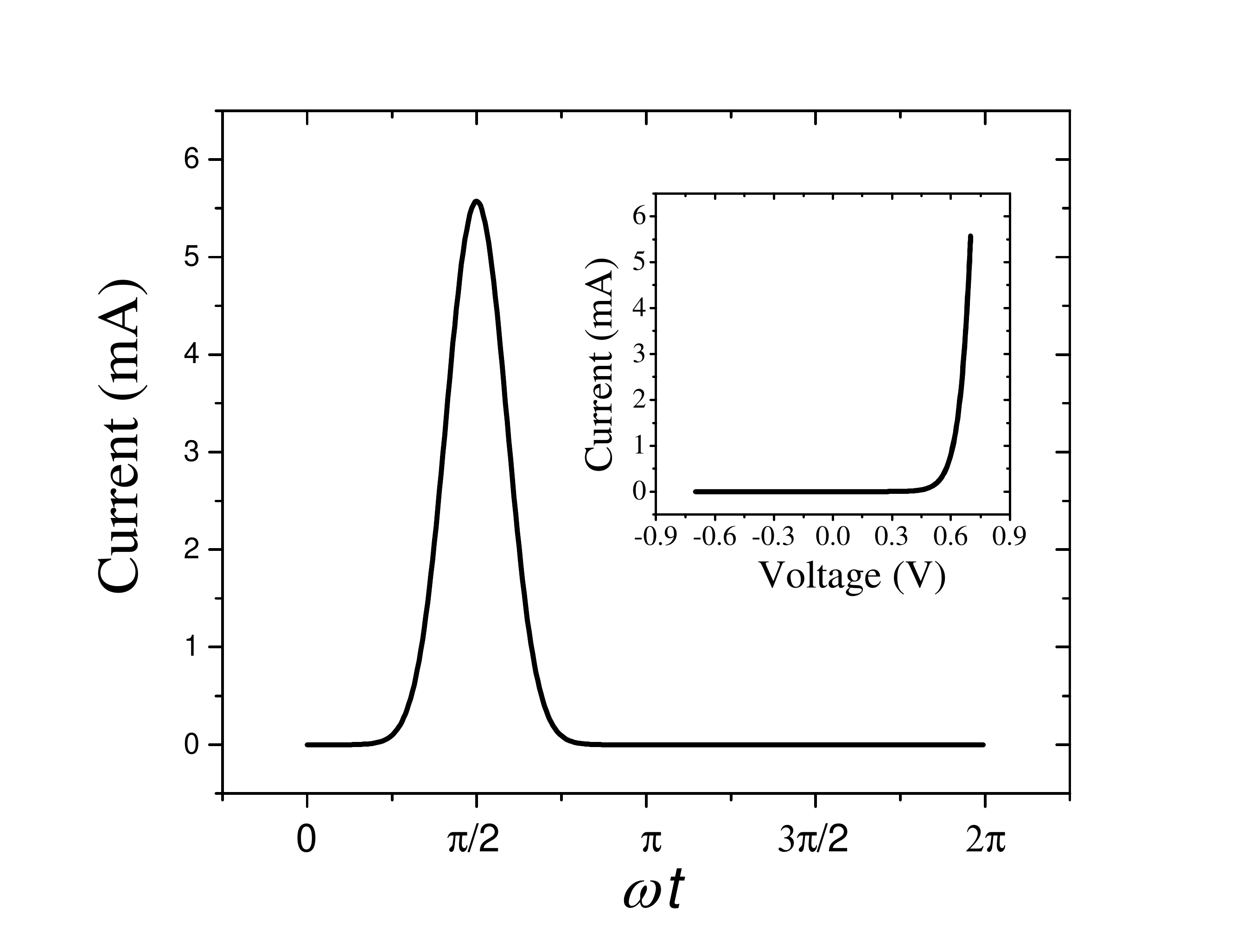}
(c)\includegraphics[angle=0,width=7cm]{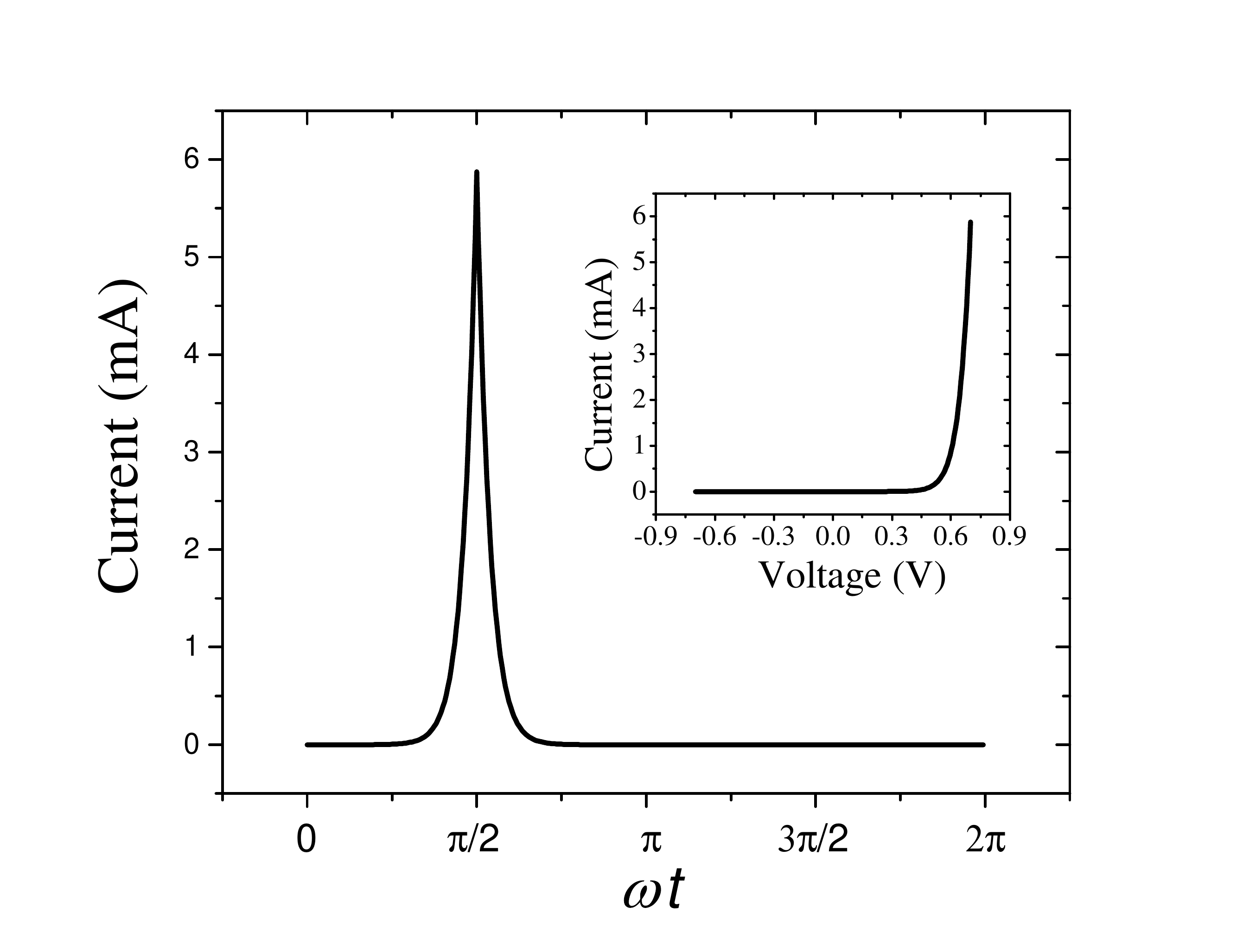}
(b)\includegraphics[angle=0,width=7cm]{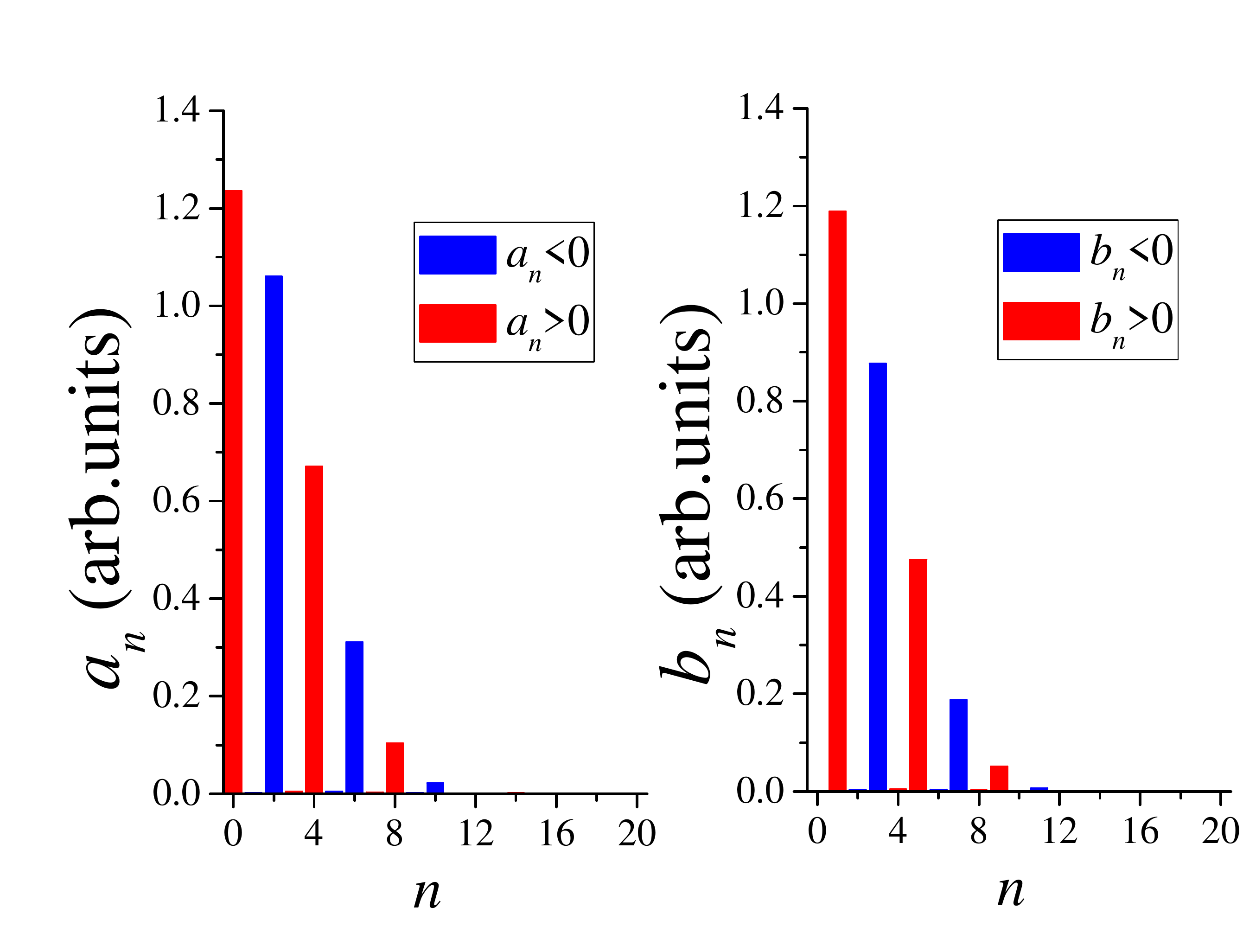}
(d)\includegraphics[angle=0,width=7cm]{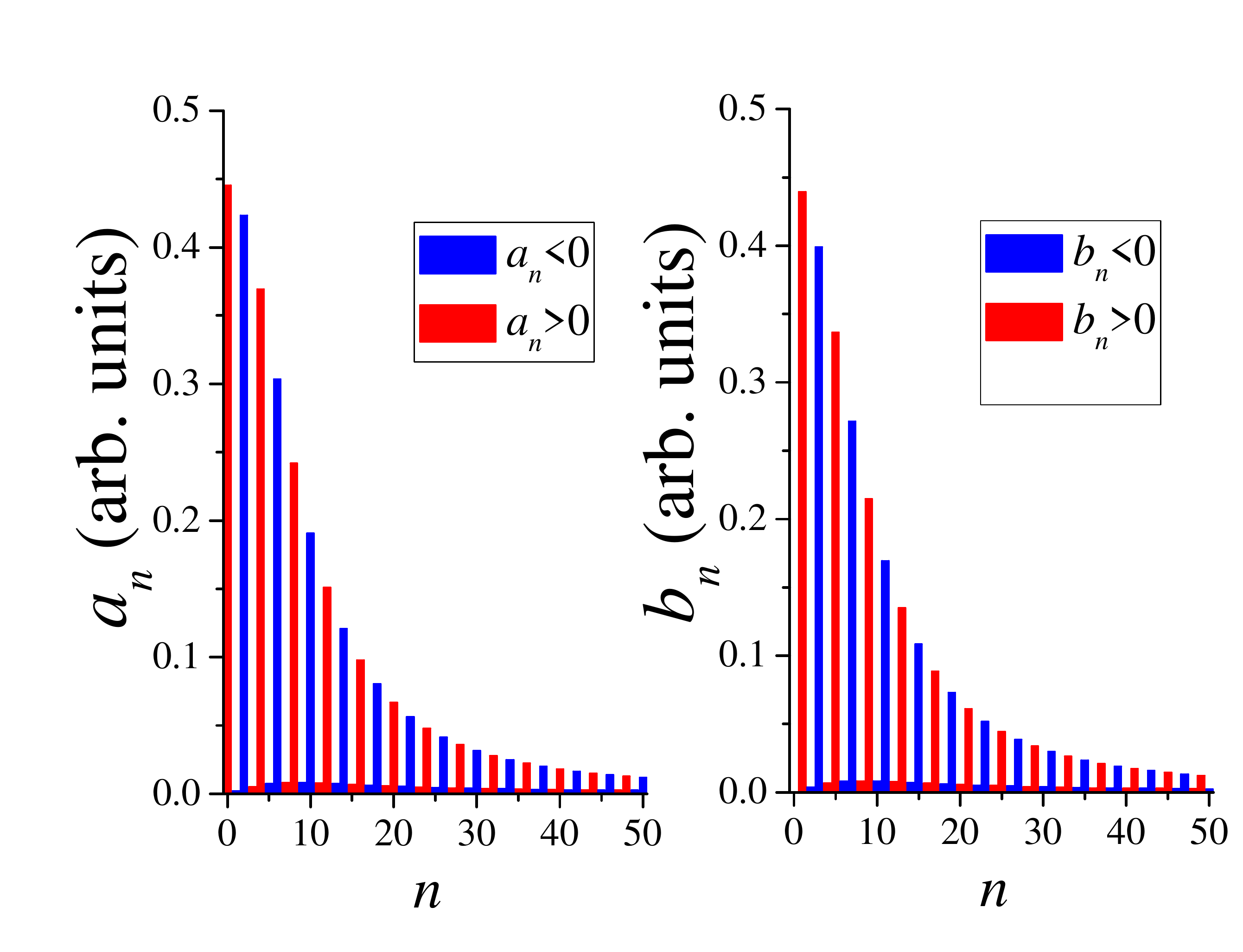}
\end{center}
\caption{A diode driven by a sinusoidal (left) and a triangular (right) voltage waveform. (a), (c) Current as a function of $\omega t$, and (b), (d) coefficients of
	the Fourier series. Insets in (a) and (c): current as a function of voltage. In (b) and (d), the magnitude (absolute value) of $a_n$-s and $b_n$-s is shown in the vertical direction, while the sign of $a_n$-s and $b_n$-s is represented by color.}\label{fig:D}
\end{figure}

We measured current-voltage curves of  resistor, diode, and memristive device, and used the $I(\omega t)$ dependence to find the coefficients $a_n$ and $b_n$ of Fourier series, Eq. (\ref{eq:1}), for each of these devices. Fig.~\ref{fig:R} presents the measurement results for the resistor. In the case of sinusoidal driving (Fig.\ref{fig:R}(a) and (b)), the result is trivial, since only $b_1\neq 0$. Higher $b_n$-s harmonics appear in the case of triangular wave driving (Fig.\ref{fig:R}(c) and (d)). In this case, we have verified that the coefficients $b_n(n)$ decay as $\sim 1/n^2$, in agreement with Feature 2 above.

Fig.~\ref{fig:D} presents the measurement results for the diode. We note that, as expected, the current-voltage curves (Fig.~\ref{fig:D}(a) and Fig.~\ref{fig:D}(c)) have a strong nonlinearity characteristic of diodes. Importantly, only even $a_n$-s and odd $b_n$-s are different from zero. Another observation is that the sign of the non-zero coefficients alternates. Moreover, the frequency content is much richer for the case of the triangular excitation, including a much slower convergence. We attempted to fit the $a_n(n)$ and $b_n(n)$ in Fig. \ref{fig:D}(b) and (d) with both the expected $1/n^2$ (according to Feature 2) and $1/n$. However, none of these fits converged to the data points. This is due to the degree of non-linearity of the diode and the amplitude of the waveform (see Eqs.~(\ref{eq:d1}) and~(\ref{eq:d2}) below).

 \begin{figure}[tb]
 \begin{center}
(a)\includegraphics[angle=0,width=7cm]{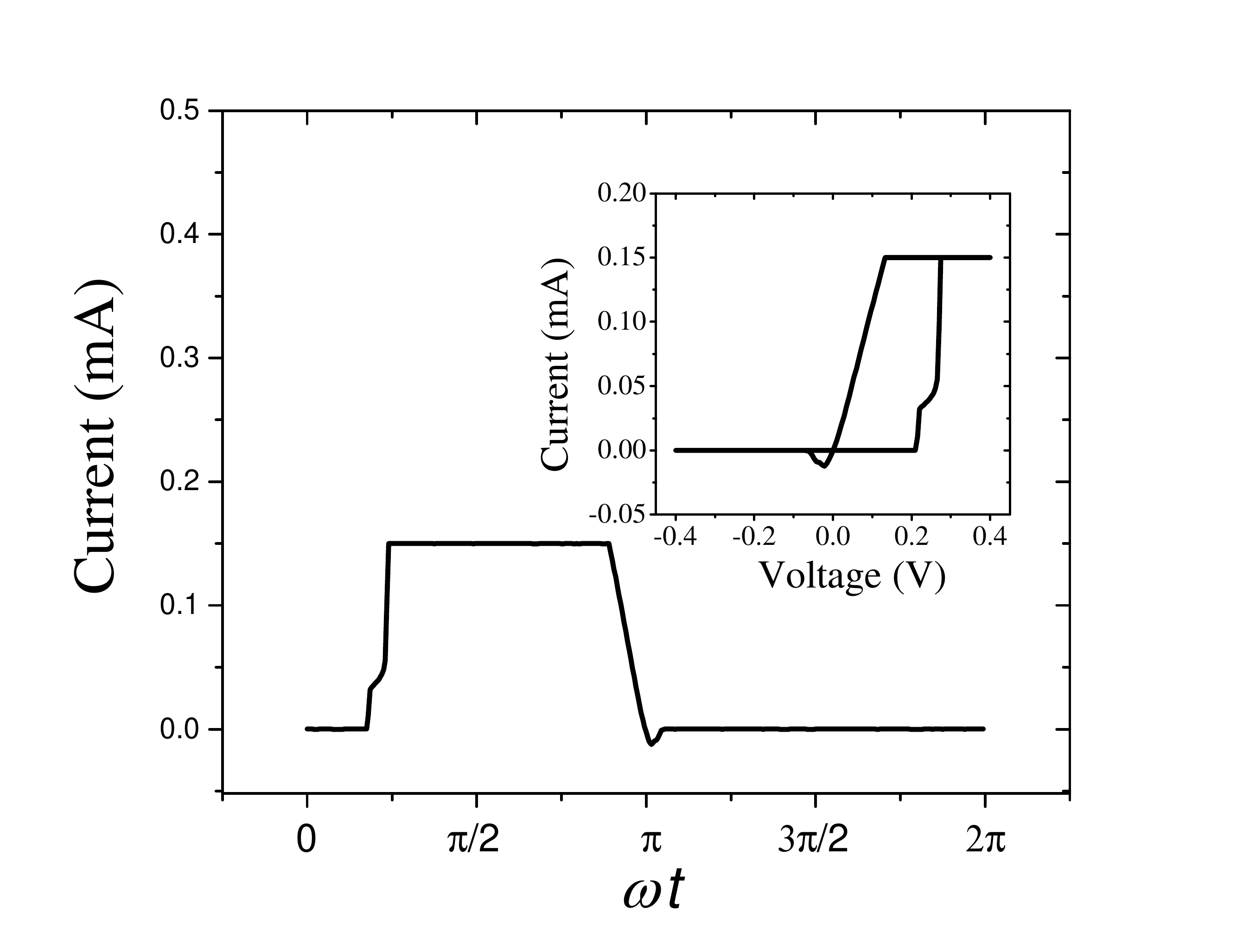}
(c)\includegraphics[angle=0,width=7cm]{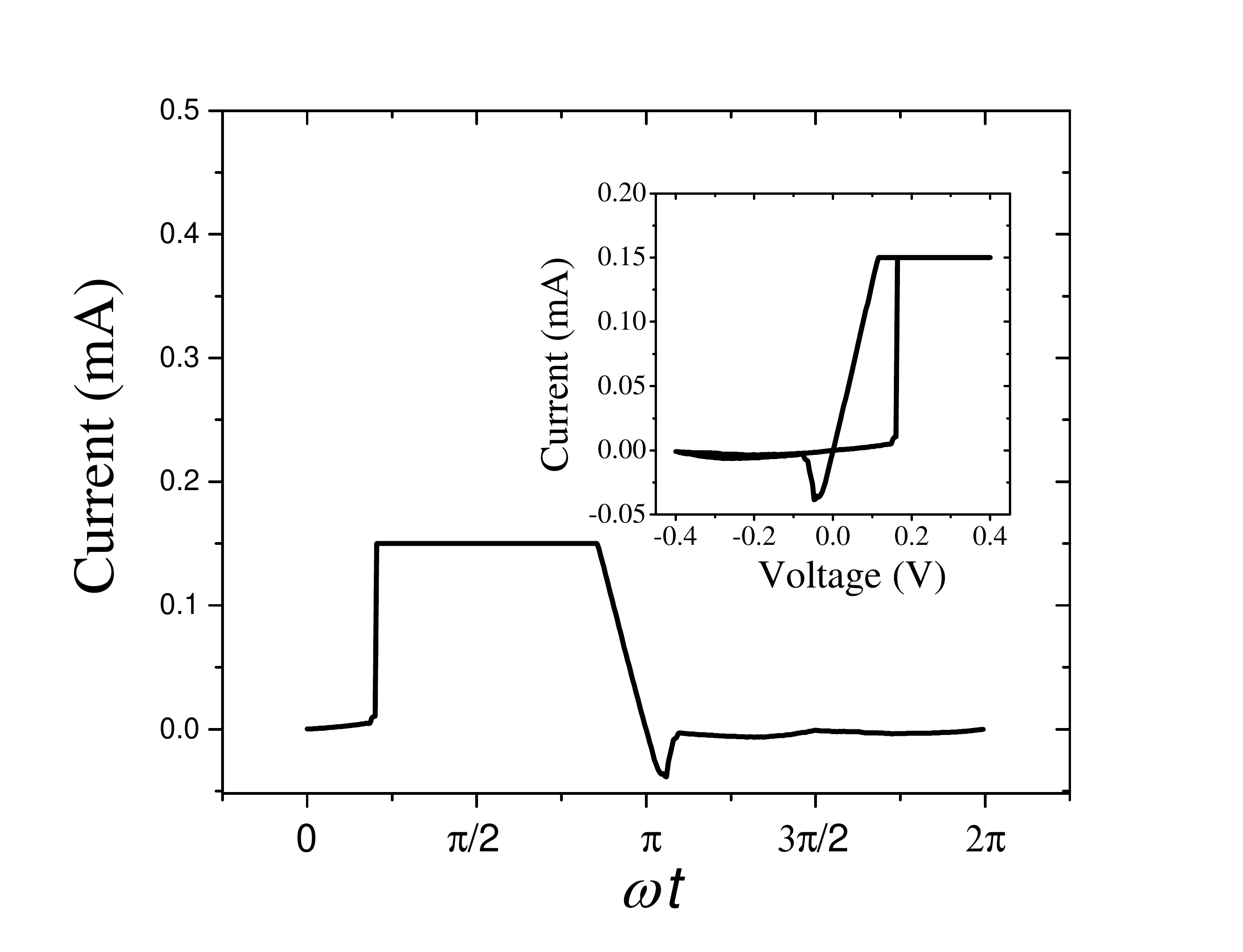}
(b)\includegraphics[angle=0,width=7cm]{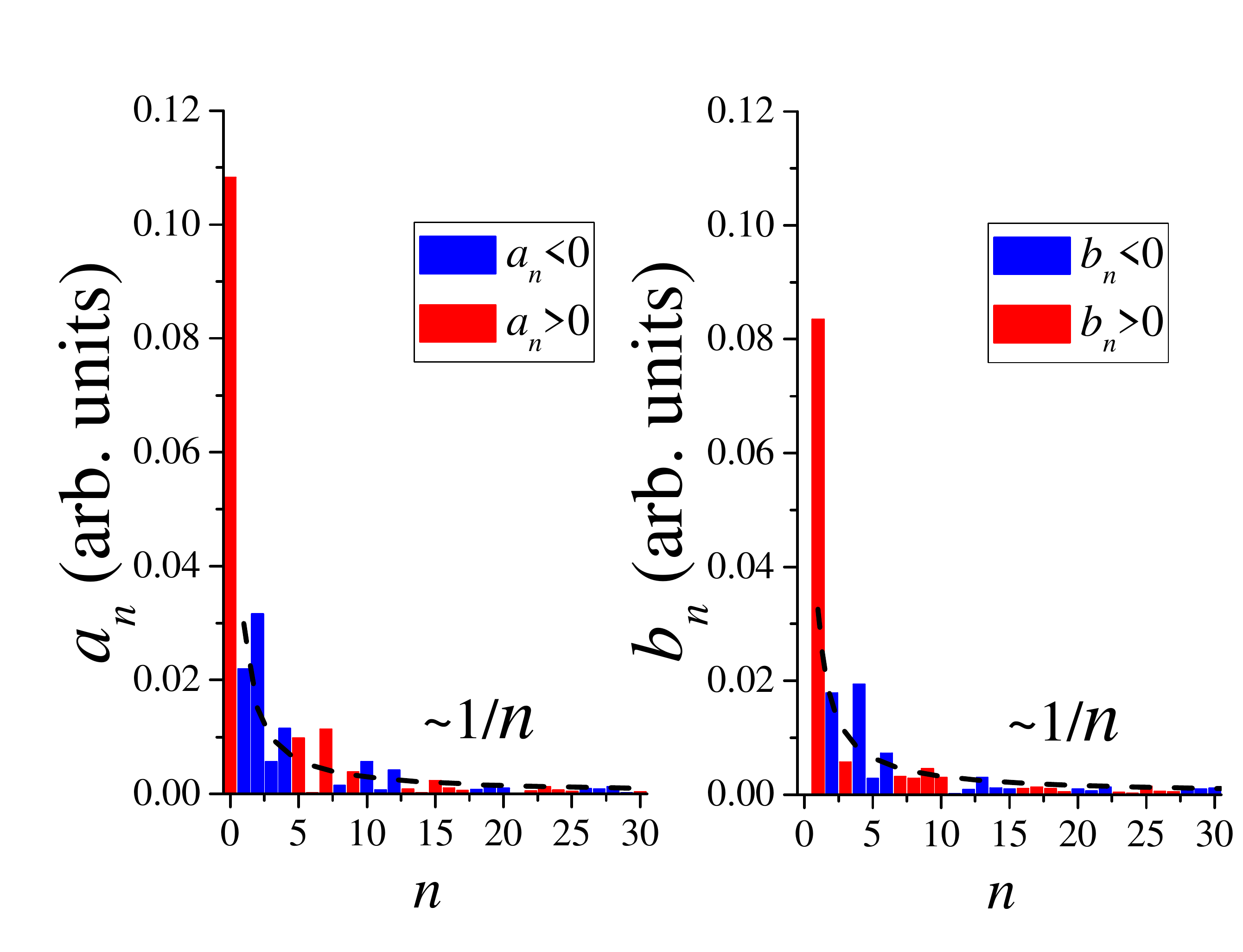}
(d)\includegraphics[angle=0,width=7cm]{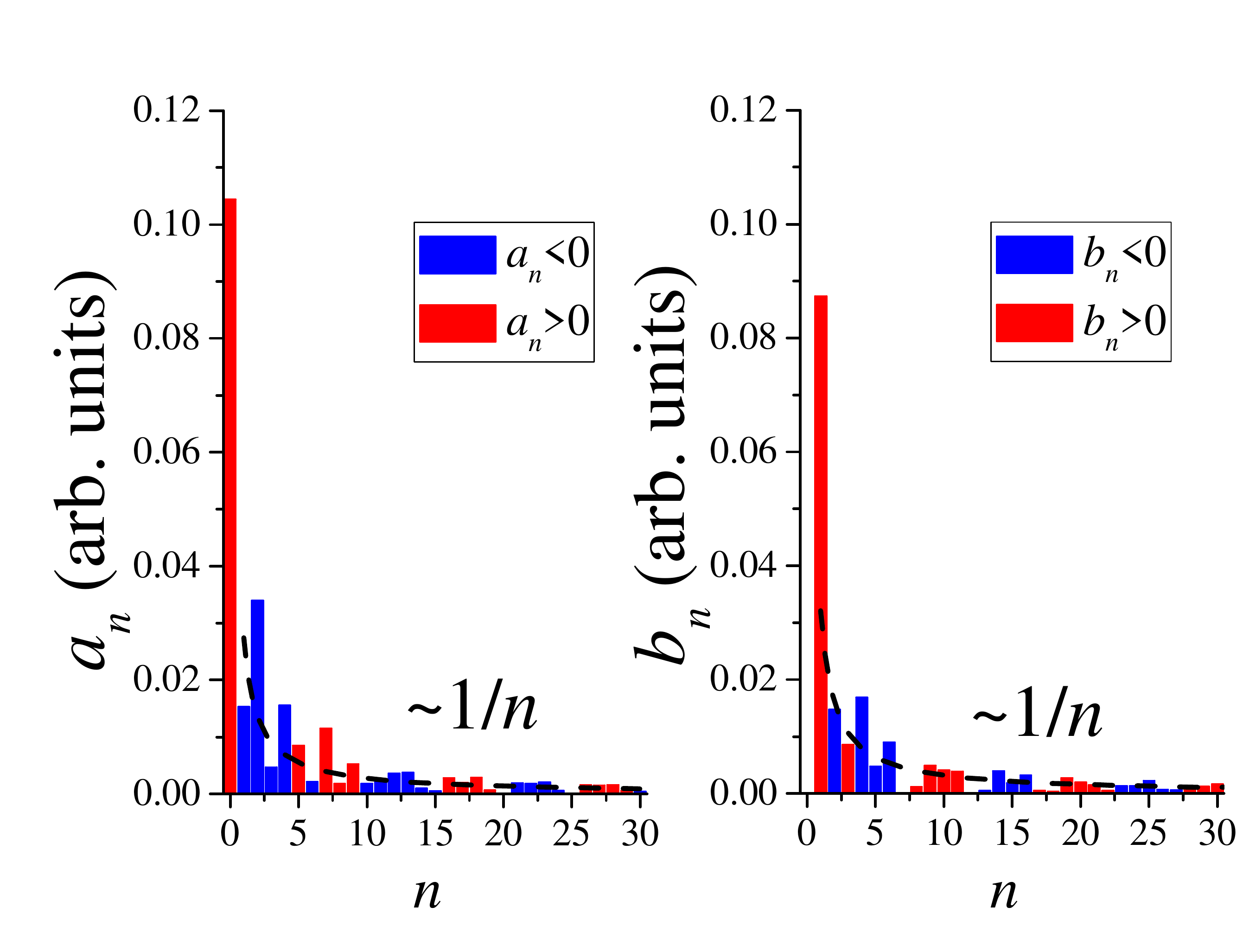}
\end{center}
\caption{Memristive device driven by sinusoidal (left) and triangular (right) voltage waveforms. (a), (c) Current as a function of $\omega t$, and (b), (d) coefficients of Fourier series. In (b) and (d) the $1/n$ fits are also shown. Insets in (a) and (c): current-voltage curve. In these measurements, 150~$\mu$A current compliance was used.}\label{fig:M}
\end{figure}

The response of the M+SDC Cr memristive cell is typical of electrochemical metallization cells~\cite{schindler2007bipolar,valov2011electrochemical}. According to the inset in Fig.~\ref{fig:M}(a) and (c), the off-to-on switching occurs at a positive threshold of about 0.2 V, while the on-to-off transition occurs at a negative threshold of a smaller magnitude. Fig. \ref{fig:M}(b) and (d) present the coefficients of the Fourier expansions.

Clearly, the first ones, $a_0$ and $b_1$, are dominant in both cases. The decay of the other coefficients occurs slower than that in the diode, and can be fitted by $1/n$,
which is consistent with Feature 1 above (the current $I(\omega t)$ has clear discontinuities, as seen in Fig.~\ref{fig:M}(a) and (c)). Moreover, the positive and negative coefficients are now grouped into small groups, and are not monotonically decreasing with $n$. A similar behavior was previously observed in numerical simulations~\cite{Joglekar12a}. Small differences between Fig.~\ref{fig:M}(a) and (c) can be explained by the well-known cycle-to-cycle variability of memristive devices.

\section{Discussion} \label{sec:discuss}

Generally, the coefficients of the Fourier series contain the entire information about the waveform they represent.
In some cases, they provide the essential information about the initial signal in a very compact way, as in the case of the resistor. Our measurements show that the Fourier content of non-linear and memristive devices can be quite broad. At the same time, some important differences induced by memory can still be recognized.

First of all, we argue that the property of $a_n=0$ for even $n$ and $b_n=0$ for odd $n$ is general for devices without memory. These conditions originate from the reflection symmetry of $I(\omega t)$ curves with respect to $\omega t=\pi/2$ and $\omega t=3\pi/2$ lines on the intervals $[0,\pi]$ and $[\pi,2\pi]$, respectively. Second, the features of the Fourier series coefficients (see below Eq. (\ref{eq:1})) can explain
the harmonic content of the  resistor and memristive device response. In the case of the diode, we have observed a significant deviation of the coefficients from $1/n$ and $1/n^2$ behavior (see below). Third, we have observed that the convergence of the Fourier coefficients is faster for a sinusoidal excitation, which makes it preferable to other waveforms, as it provides the most compact signal representation.

Moreover, for resistive devices in general, the requirement of $I=0$ at $V=0$ leads to some constraints on $a_n$-s. Specifically, writing Eq. (\ref{eq:1}) for $\omega t=0$ and $\omega t=\pi$ we get
\begin{equation}\label{eq:2}
  \frac{a_0}{2} +\sum\limits_{n=1}^\infty a_n=0.
\end{equation}
and
\begin{equation}\label{eq:2}
  \frac{a_0}{2} +\sum\limits_{n=1}^\infty \left(
  -1\right)^{n}a_n=0,
\end{equation}
respectively.
Clearly, the point $\omega t=2\pi$ does not lead to any new relation.

In addition, we note that the memristive devices that were used in our experiments are bipolar memristive devices, which as the most common ones. Another large class of memristive devices are unipolar memristive devices (the thermistor is a notable example)~\cite{diventra09a}. The response of unipolar memristive devices (when they driven by sinusoidal or triangular pulses) satisfies $I(\omega t+\pi)=- I(\omega t)$. It is evident that the even harmonics are missing in the Fourier series of unipolar memristive devices. This implies that all cosine terms are zero in unipolar devices without memory (for example, in resistors, see Fig.~\ref{fig:R}(b) and (d)).

\begin{table}[tb]
  \begin{center}
    \begin{tabular}{c|c|c} 
      Device & Distinctive Fourier features & Rationale\\
      \hline \hline
      Linear resistors & All coefficients are zero except $b_1$ & Fig.
      \ref{fig:R}(b)\\
       \hline
      Non-linear devices without & All odd $a_n$-s and even $b_n$-s are zero & Fig.
      \ref{fig:D}(b)\\
      memory (diodes, etc.) & & \\
       \hline
      Binary/multi-state memristive elements, & Slow converging series ($\sim 1/n$)& Fig.
      \ref{fig:M}(b) and Ref.~\cite{Joglekar12a}
       \\
      slow-driven analog memristive elements & & \\
      \hline
      Fast-driven analog memristive elements & Fast converging series & Ref.~\cite{Joglekar12a} \\
      & ($\sim 1/n^2$ or faster) & \\
      \hline
      Unipolar memristive elements & All $a_n$-s and even $b_n$-s are zero & see Sec.~\ref{sec:discuss} \\
      \hline
    \end{tabular}
  \end{center}
  \caption{Summary of experimental results for (sinusoidally-driven) systems with and without memory. }\label{tab:table1}
\end{table}

To better understand the deviation of the diode's coefficients from the characteristic features of Fourier expansions~\cite{arfken1999mathematical}, let us first derive these coefficients for the case of triangular driving.
Using the Shokley diode model~\cite{shockley1949theory}
\begin{equation}\label{eq:diode}
  I=I_s\left(e^{\alpha V}-1 \right)
\end{equation}
for $V>0$, and approximating the current by $I=0$ for $V<0$, it is not difficult to derive the following expressions for the Fourier series coefficients:
\begin{equation}\label{eq:d1}
a_n=\frac{2I_sk\alpha\cos\frac{n\pi}{2}}{\pi}\;\frac{e^{\frac{k\pi\alpha}{2}}-\cos\frac{\pi n}{2}}{n^2+k^2\alpha^2}
\end{equation}
and
\begin{equation}\label{eq:d2}
b_n=\frac{2I_sk\alpha\sin\frac{n\pi}{2}}{\pi}\;\frac{e^{\frac{k\pi\alpha}{2}}-\cos\frac{n\pi}{2}}{n^2+k^2\alpha^2}.
\end{equation}
where $k=V_0/\left(\pi/2\right)$, and $V_0$ is the waveform amplitude. The deviation of the diode's coefficient behavior from the expected $1/n^2$ trend (according to Feature 2) can be explained by the importance of the $k^2\alpha^2$ term compared to $n^2$ for small values of $n$.

\begin{figure}[tb]
 \begin{center}
\includegraphics[angle=0,width=7cm]{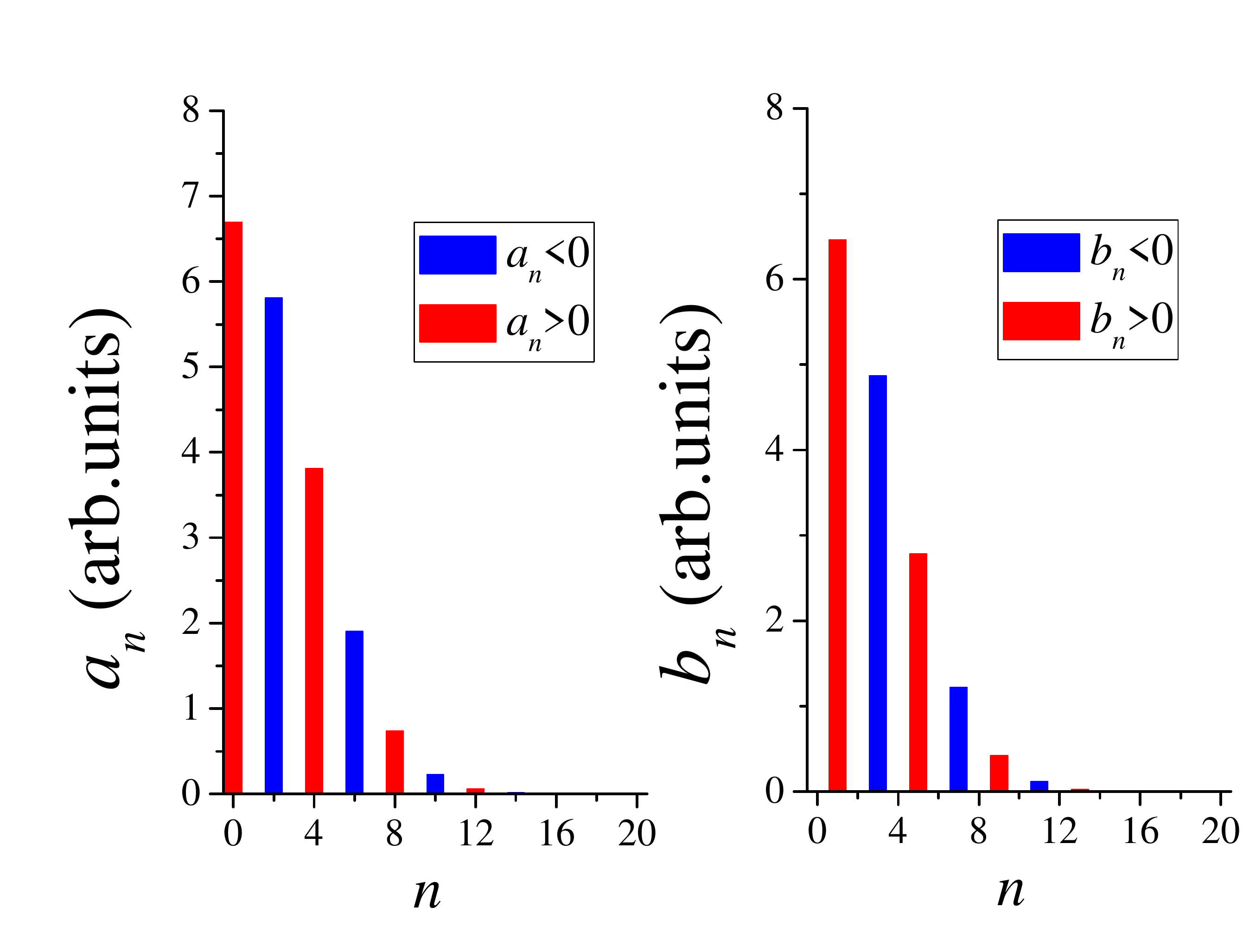}
\end{center}
\caption{Fourier coefficients for a sinusoidally-driven diode plotted using  Eqs.~(\ref{eq:d1sin}) and (\ref{eq:d2sin}).}\label{fig:DiodeModel}
\end{figure}

In the case of the sinusoidal driving voltage, the coefficients are given by
\begin{equation}\label{eq:d1sin}
a_n=2I_s\cos\frac{n\pi}{2}I_n\left(\alpha V_0 \right)
\end{equation}
and
\begin{equation}\label{eq:d2sin}
b_n=2I_s\sin\frac{n\pi}{2}I_n\left(\alpha V_0 \right),
\end{equation}
where $I_n(..)$ is the modified Bessel function of the first kind. Fig.~\ref{fig:DiodeModel} shows the Fourier coefficients Eqs.~(\ref{eq:d1sin}) and (\ref{eq:d2sin}) found using $I_s=14.63$~nA, $\alpha=20.84$~V$^{-1}$ extracted from Ref.~\cite{cataldo2016}, and $0.7$~V sinusoidal voltage amplitude. This plot almost identically repeats the pattern of experimentally measured Fourier spectrum of the diode (Fig.~\ref{fig:D}(b)).

Next, we introduce a qualitative characteristic of memory -- {\it the measure of hysteresis}. For this purpose, we consider the collection of Fourier coefficients $a_n$ and $b_n$ as components of a vector, and use the squared length of the vector built out of Fourier components not present in the non-linear devices (odd $a_n$-s and even $b_n$-s). We normalize this quantity by the squared length of the vector composed of all Fourier components. The measure of hysteresis we define is then:
\begin{equation}\label{eq:measure}
\kappa=\frac{\sum\limits_{n=0}^\infty\left( a_{2n+1}^2+b_{2n+2}^2\right)}{\sum\limits_{n=0}^\infty\left( a_{n}^2+b_{n+1}^2\right)}.
\end{equation}

\begin{figure}[tb]
 \begin{center}
(a)\includegraphics[angle=0,width=7cm]{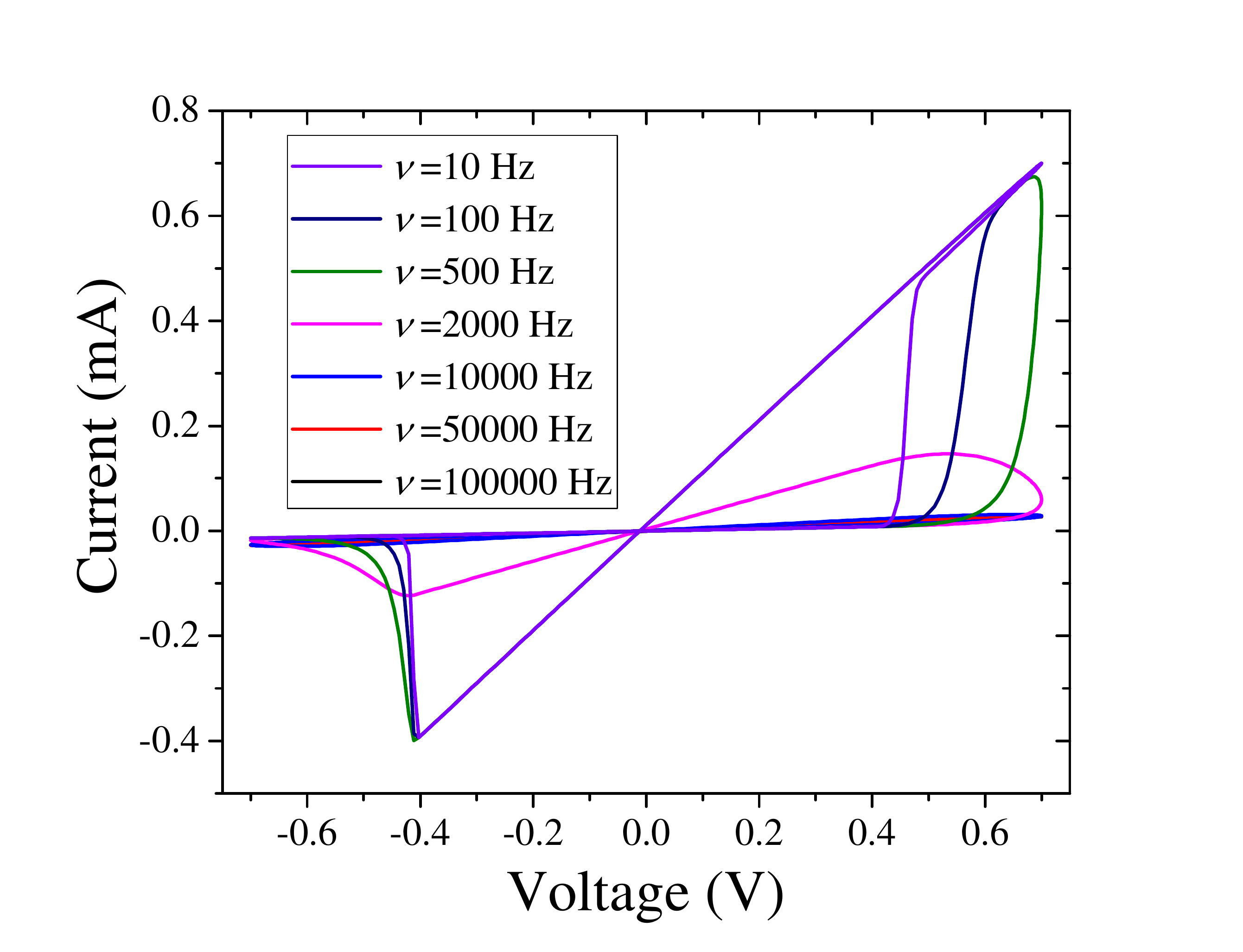}
(b)\includegraphics[angle=0,width=7cm]{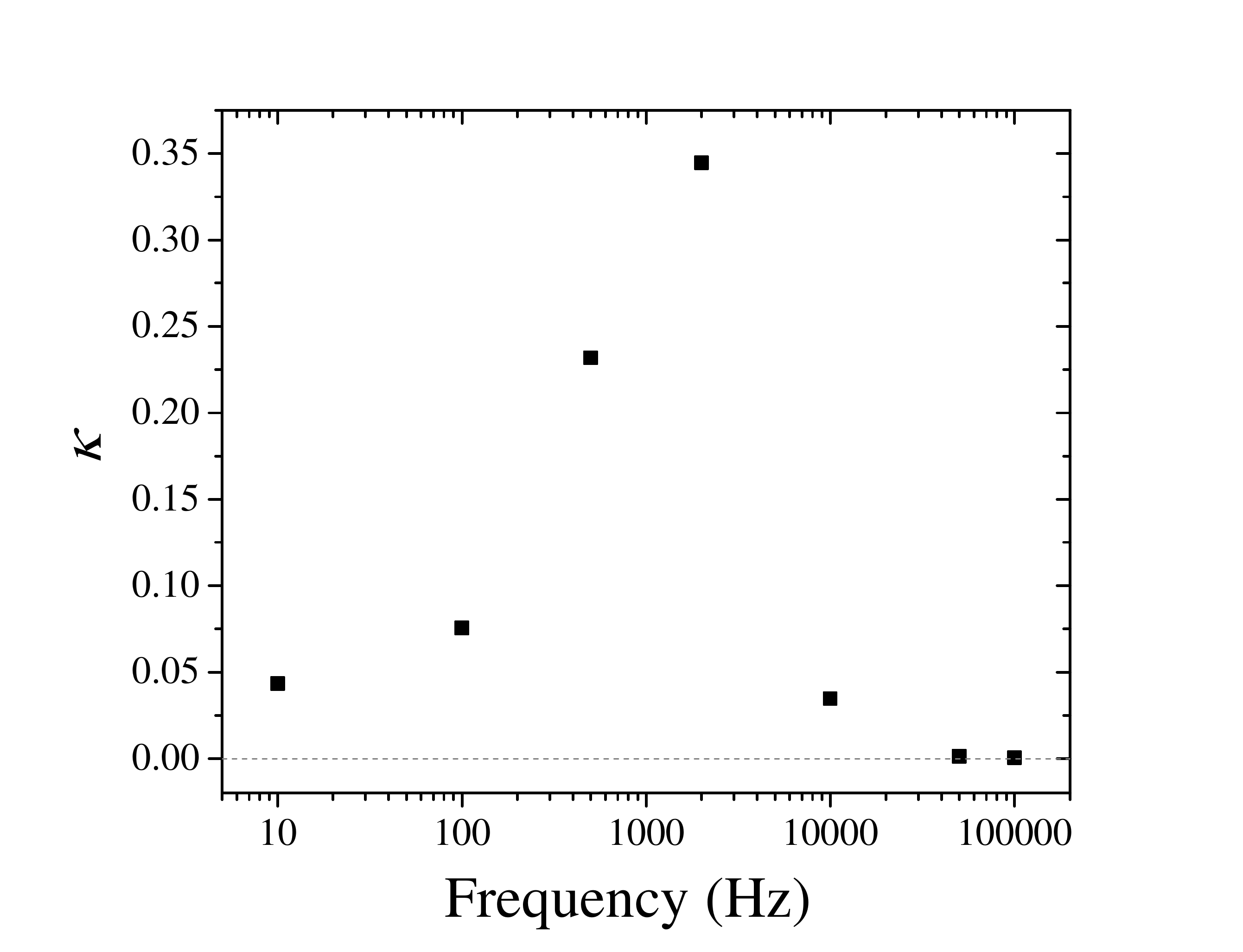}
\end{center}
\caption{(a) Current-voltage curves plotted with the use of Eq.~(\ref{eq:model1}) memristive model with $V(t)=V_0\sin (2\pi\nu t)$ for $V_0=0.7$~V.
(b) Measure of hysteresis as a function of frequency calculated for the curves in (a). These plots were obtained using the following parameter values: $R_{on}=1$~k$\Omega$, $R_{off}=50$~k$\Omega$, $V_t=0.4$~V, $g=10^5$ $\Omega$V$^{-1}$s$^{-1}$. }\label{fig:measure}
\end{figure}

To illustrate the definition of $\kappa$, we consider the model of a first-order voltage-controlled memristive system with threshold
\begin{eqnarray}
 V(t) &=& R(x)I(t), \;\; R=R_{on}x+R_{off}(1-x) ,\label{eq:model1} \\
\frac{\textnormal{d}x}{\textnormal{d}t}&=&gf_w(x,V)\cdot\left\{
\begin{array}{ll}
                   (V-V_t),\; V>V_t\\
                   (V+V_t),\; V< -V_t \\
                   0,\;\;\textnormal{otherwise}
                \end{array}
\right.
, \label{eq:model2}
\end{eqnarray}
where $x$ is the internal state variable, $R_{on}$ and $R_{off}$ are resistances in the on and off states, respectively, $g$ is a constant, $V_t>0$ is the voltage threshold, and $f_w(x,V)$ is the window function defined by
\begin{equation}
 f_{w} (x,V)= \left\{
\begin{array}{ll}
                  1-x,\; V\geq 0 \\
                  x,\;\;\;\;\; V < 0
                \end{array}
\right. \;\;  .\label{eq:wf}
\end{equation}
Fig.~\ref{fig:measure}(a) presents the current-voltage curves for the above model that have all typical characteristics of memristive systems~\cite{chua76a}. At high frequencies, the hysteresis is strongly reduced as the internal state does not have enough time to follow the time-dependent voltage. The largest hysteresis is observed at an intermediate frequency. The frequency dependence of $\kappa$ in Fig.~\ref{fig:measure}(b) indeed qualitatively reproduces the frequency behavior of the model.

At the same time, we note that the maximum of $\kappa$ corresponds to the $\nu=2000$~Hz curve, whose hysteresis is not the largest. This is due to the fact that the size of the hysteresis is not the only factor that contributes to $\kappa$, but also  other factors such as the specifics of the input shape. In any case, $\kappa$ can be employed as a useful parameter for the characterization of hysteresis curves.

After showing the potential of $\kappa$ to quantify the memory content in the current-voltage response, we apply this measure to our experimental results. For the sinusoidally-driven resistor, diode, and memristor we find $\kappa=4.9\cdot 10^{-9}$, $2.0\cdot 10^{-5}$, and $0.072$, respectively. In the case of triangular voltage waveform, the values are $9.3\cdot 10^{-5}$, $0.00091$, and $0.055$, respectively. The data clearly indicate that the measure of hysteresis $\kappa$ can be used to distinguish the response of a memory device (large/significant $\kappa$) from that of a resistor and a diode (small/insignificant $\kappa$). It is also interesting that the low-frequency value of $\kappa$ in Fig.~\ref{fig:measure} is close to the experimentally found values with the memristive devices (related to the low-frequency limit as well).

\section{Conclusion}

We have shown experimentally that resistive devices with memory can be easily distinguished from devices without memory by the analysis of their Fourier series coefficients. Moreover, we have demonstrated that the waveform of the driving signal has an important effect on the Fourier spectrum. Our main observations are summarized in Table \ref{tab:table1} for the case of sinusoidal driving. The last line of this table is formulated based on Ref.~\cite{Joglekar12a}, which is also supported by the general tendency of smoother $I(\omega t)$ dependence at higher frequencies (because of the
reduction of the memristive hysteresis~\cite{pershin11a}). We believe this work gives further support to the notion that Fourier analysis is a powerful tool to distinguish the response of systems with memory from those without. We also expect that the Fourier analysis may be useful in the studies of preconditioning effects~\cite{Gupta_2020} and cycle-to-cycle variability effects.

\section*{Acknowledgments}

M.D. acknowledges support the Center for Memory and Recording Research at the University of California, San Diego, and DARPA under grant No. HR00111990069.

\bigskip
\bigskip
{\bf REFERENCES}
\bigskip
\bigskip

\bibliographystyle{unsrt}
\bibliography{memcapacitor}

\begin{thebibliography}{10}

\bibitem{chua76a}
Leon~O. Chua and Sung~Mo Kang.
\newblock Memristive devices and systems.
\newblock {\em Proc. {IEEE}}, 64:209--223, 1976.

\bibitem{Joglekar12a}
Y.~N. {Joglekar} and N.~{Meijome}.
\newblock Fourier response of a memristor: Generation of high harmonics with
  increasing weights.
\newblock {\em IEEE Transactions on Circuits and Systems II: Express Briefs},
  59(11):830--834, 2012.

\bibitem{Cohen12a}
Guy~Z. Cohen, Yuriy~V. Pershin, and Massimiliano Di~Ventra.
\newblock Second and higher harmonics generation with memristive systems.
\newblock {\em Applied Physics Letters}, 100(13):133109, 2012.

\bibitem{Biolek14a}
D.~{Biolek}, Z.~{Biolek}, V.~{Biolkova}, and Z.~{Kolka}.
\newblock Some regularities of the spectral content of the responses of
  memristive systems to sinusoidal excitation.
\newblock In {\em 2014 European Modelling Symposium}, pages 473--478, 2014.

\bibitem{Hu19a}
W.~{Hu} and R.~{Wei}.
\newblock An analytic approach to nonlinearity analysis of memristor.
\newblock {\em IEEE Transactions on Electron Devices}, 66(6):2589--2594, 2019.

\bibitem{knowm}
Self directed channel memristors.
\newblock \url{https://knowm.org/downloads/Knowm_Memristors.pdf}.
\newblock Accessed: 2020-03-20.

\bibitem{arfken1999mathematical}
George~B Arfken and Hans~J Weber.
\newblock Mathematical methods for physicists, 1999.

\bibitem{raisbeck1955order}
Gordon Raisbeck.
\newblock The order of magnitude of the fourier coefficients in functions
  having isolated singularities.
\newblock {\em The American Mathematical Monthly}, 62(3):149--154, 1955.

\bibitem{schindler2007bipolar}
Christina Schindler, Sarath Chandran~Puthen Thermadam, Rainer Waser, and
  Michael~N Kozicki.
\newblock Bipolar and unipolar resistive switching in {C}u-doped {SiO}$_2$.
\newblock {\em IEEE Transactions on Electron Devices}, 54(10):2762--2768, 2007.

\bibitem{valov2011electrochemical}
Ilia Valov, Rainer Waser, John~R Jameson, and Michael~N Kozicki.
\newblock Electrochemical metallization memories—fundamentals, applications,
  prospects.
\newblock {\em Nanotechnology}, 22(25):254003, 2011.

\bibitem{diventra09a}
Massimiliano {Di Ventra}, Yuriy~V. Pershin, and Leon~O. Chua.
\newblock Circuit elements with memory: Memristors, memcapacitors, and
  meminductors.
\newblock {\em Proc. {IEEE}}, 97(10):1717--1724, 2009.

\bibitem{shockley1949theory}
William Shockley.
\newblock The theory of p-n junctions in semiconductors and p-n junction
  transistors.
\newblock {\em Bell System Technical Journal}, 28(3):435--489, 1949.

\bibitem{cataldo2016}
Enrico Cataldo, Alberto Di~Lieto, Francesco Maccarrone, and Giampiero Paffuti.
\newblock Measurements and analysis of current-voltage characteristic of a pn
  diode for an undergraduate physics laboratory.
\newblock {\em arXiv preprint arXiv:1608.05638}, 2016.

\bibitem{pershin11a}
Yuriy~V. Pershin and Massimiliano Di~Ventra.
\newblock Memory effects in complex materials and nanoscale systems.
\newblock {\em Advances in Physics}, 60:145--227, 2011.

\bibitem{Gupta_2020}
Vishal Gupta, Giulia Lucarelli, Sergio Castro-Hermosa, Thomas Brown, and Marco
  Ottavi.
\newblock Investigation of hysteresis in hole transport layer free metal halide
  perovskites cells under dark conditions.
\newblock {\em Nanotechnology}, 31(44):445201, 2020.

\end{thebibliography}

\end{document}